# Open Access does not increase citations for research articles from The Astrophysical Journal

Michael J. Kurtz and Edwin A. Henneken
Harvard-Smithsonian Center for Astrophysics

0. ABSTRACT

We demonstrate conclusively that there is no "Open Access Advantage" for papers from the Astrophysical Journal. The two to one citation advantage enjoyed by papers deposited in the arXiv e-print server is due entirely to the nature and timing of the deposited papers. This may have implications for other disciplines.

1. INTRODUCTION

Lawrence (2001) first showed that research articles which have been posted on the internet have higher citation rates than articles which have not. The initial assumption was that this was causal, because more people could access and read these papers they were cited more. This assumption was strengthened by the analysis of Harnad and Brody (2004), who used physics articles which were either deposited in arXiv (Ginsparg, 1996), or not, and found that arXiv deposited articles were better cited by more than a factor of two.

Kurtz, et al (2005a) investigated three possible causes for the effect: Early Access (EA), arXiv deposited papers are cited more because they are available several months before the journal versions; Quality Bias (QB), either the best researchers tend to use arXiv, or researchers tend to post their best papers; and Open Access (OA), by being available for free on the internet more people are able to read the arXiv deposited papers, thus they are more cited.

Kurtz, et al (2005a) found that the EA and QB postulates were able to account for all their measurements, which were for papers in astrophysics. They were unable to find any OA effect. They explained this by suggesting that in a well funded field like astrophysics essentially everyone who is in a position to write research articles has full access to the literature.

Using different methodologies Moed (2007) studied the literature of solid state physics and came to very similar conclusions. The Southampton group responded by refining their analysis (Brody, et al 2007), making plausible arguments explaining Kurtz, et al. and Moed's results, and reasserting their claims that OA is the dominant effect.

The difficulty with distinguishing among the various hypotheses is that, until now, there has been no case where they are clearly separated. Just the act of an author self depositing a paper on line in an open archive creates a possible bias compared to the author/paper which is not deposited.

We have found a dataset which does not have this bias; in 1997 the on-line Astrophysical Journal was a fully open access journal, no subscription barrier existed. One was erected on January 1, 1998. By comparing the citation experiences of different sets of articles from the 1997 and 1998 ApJ we are able to show conclusively the (non) existence of the OA effect.

2. DATA

We use citation records for papers published in the Astrophysical Journal in 1997 and 1998 taken from the citation database of the Smithsonian/NASA Astrophysics Data System (Kurtz, et al 2005b; ADS). The data are divided into four categories, by year, and by whether they were deposited in the arXiv e-print server (Ginsparg, 1996). The matching of journal articles with their arXiv versions is a routine part of the ongoing arXiv-ADS collaboration (Henneken, et al, 2006a). In 1997 there were 965 ApJ papers deposited in arXiv, and 1408 which were not; in 1998 1243 were deposited and 1105 were not.

The Astrophysical Journal is among the first major scholarly journals to have a full online version; the Letters section went online (HTML and PDF) on July 1, 1995, with the main journal coming online on January 1, 1997 (Boyce, et al 1996). By mid 1997 the online version was being read more via the ADS than the print version was in all the world's research libraries, combined (Kurtz, et al 2000).

The Astrophysical Journal implemented access controls, requiring potential readers to have subscriptions, on January 1, 1998; from that time on they have had a three year moving window, so articles can be viewed only by subscribers until they are 36 months old. The pre-1998 articles were not put behind the subscription wall when it was erected, thus they have always been "open access."

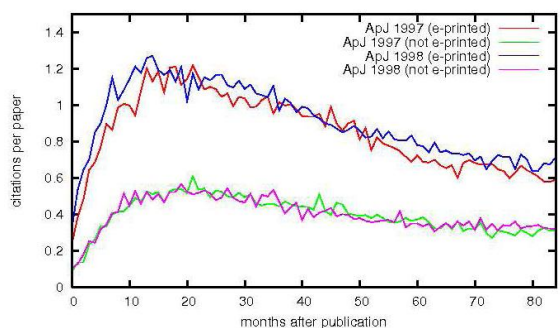

## 3. ANALYSIS

The figure plots the citation histories for the four subsets (1997 deposited in arXiv, or not and 1998 deposited in arXiv, or not) of the Astrophysical Journal papers. The most obvious effect (Henneken, et al 2006b) is that arXiv deposited papers are cited at about twice the rate of non-deposited papers; next we see that the 1998 arXiv deposited papers have their peak citation rate earlier than the 1997 deposited papers, part of a long term trend shown by Brody, et al. (2006).

Here we will discuss the bottom two lines on the graph, representing the 1997 and 1998 non-arXiv deposited ApJ articles. These two subsets represent a near perfect sample to compare the effects of open access absent the complicating factor of any possible self archiving bias. Both sets of articles come from authors who chose not to self archive in arXiv, but one of the sets (1997) is fully open access, while the other (1998) is fully toll or closed access.

As is clear by inspection of the figure the two curves overlay each other nearly perfectly. There is no significant difference in the citation histories of the two sets, in particular there is no significant difference for the first 36 months following publication, where the 1997 articles were OA and the 1998 articles were not. There is also no significant difference in the integral counts for these two sets of articles over any time period.

This shows directly that there is no open access citation advantage, independent of the EA and SB effects, for the Astrophysical Journal.

## 4. DISCUSSION

The existence of a large differential between articles put on-line through author self archiving, shown first by Lawrence (2001) has been repeatedly confirmed (including in our figure), and is firmly established. The issue, as posed by Kurtz, et al. (2005a) is whether all or part of this differential is caused by the differing cost components, that self archived articles are available free, without subscription, while non self archived articles require that a fee (subscription) be paid.

Here we have shown conclusively that for the Astrophysical Journal there is no cost component to the citation differential, confirming out previous result in astrophysics (Kurtz, et al. 2005a) and Moed's (2007) in condensed matter physics. There are a number of excellent arguments in favor of changing the scientific publication system to an open access model. The open access citation advantage is not one of them.